\documentclass[twocolumn,floats,showpacs,amsmath,amssymb,pre]{revtex4}
\usepackage{graphicx}
\begin{document}

\title{Antiferromagnetic Ising model in small-world networks}
\author{Carlos P. Herrero}
\affiliation{Instituto de Ciencia de Materiales de Madrid,
         Consejo Superior de Investigaciones Cient\'{\i}ficas (CSIC),
         Campus de Cantoblanco, 28049 Madrid, Spain }
\date{\today}

\begin{abstract}
The antiferromagnetic Ising model in small-world networks generated 
from two-dimensional regular lattices has been studied.
The disorder introduced by long-range connections causes frustration, 
which gives rise to a spin-glass phase at low temperature.
Monte Carlo simulations have been carried out to study the paramagnetic
to spin-glass transition, as a function of the rewiring probability $p$,
which measures the disorder strength.  The transition temperature 
$T_c$ goes down for increasing disorder, and saturates to a value 
$T_c \approx 1.7 J$ for $p > 0.4$, $J$ being the antiferromagnetic coupling.
For small $p$ and at low temperature, the energy increases linearly with
$p$. In the strong-disorder limit $p \to 1$, this model is equivalent 
to a short-range $\pm J$ spin glass in random networks.
\end{abstract}

\pacs{64.60.De, 05.50.+q, 75.10.Nr, 89.75.Hc}

% 64.60.De Statistical mechanics of model systems (Ising model, Potts model,
%          field-theory models, Monte Carlo techniques, etc)
% 05.50.+q Lattice theory and statistics (Ising, Potts, etc.)
% 75.10.Nr Spin-glass and other random models
% 89.75.Hc Networks and genealogical trees

\maketitle

\section{Introduction}

In the last few years, there has been a surge of interest in modeling
complex systems as networks or graphs, with 
nodes representing typical system units and edges playing the
role of interactions between connected pairs of units.
Thus, complex networks have been used to model several types of real-life
systems (social, economic, biological, technological), and to study various
processes taking place on them \cite{al02,do03a,ne03,ne06,co07}.
In this context, some models of networks have been designed to explain
empirical data in several fields,
as is the case of the so-called small-world networks, introduced by 
Watts and Strogatz in 1998 \cite{wa98}. 

These small-world networks are well suited to study
systems with underlying topological structure ranging from
regular lattices to random
graphs \cite{bo98,ca00}, by changing a single parameter \cite{wa99}.
They are based on a regular lattice,
in which a fraction $p$ of the links between nearest-neighbor sites
are replaced by new random connections, creating long-range
``shortcuts'' \cite{wa98,wa99}.  In the networks so generated 
one has at the same time a local neighborhood
(as in regular lattices) and some global properties of random graphs,
such as a small average topological distance between pairs of nodes.
These networks are suitable to study different kinds of 
physical systems, as neural networks \cite{la00} and man-made
communication and transportation systems \cite{wa98,la01a,ne00a}.
The importance of a short global length scale has been emphasized for 
several statistical physical problems on small-world networks.
Among these problems, one finds 
the spread of infections \cite{ku01,mo00a},
signal propagation \cite{wa98,he02b,mo04}, 
random spreading of information \cite{pa01,la01b,he03,he07,ca07}, 
as well as site and bond percolation \cite{mo00a,ne99,mo00b}.
                    
Cooperative phenomena in this kind of networks are expected to display 
unusual characteristics, associated to their peculiar topology 
\cite{ba00,sv02,ca06,do07}. 
Thus, a paramagnetic-ferromagnetic phase transition of mean-field
type was found for the Ising model on small-world networks derived 
from one-dimensional (1D) lattices \cite{ba00,gi00,lo04}.
This phase transition 
occurs for any value of the rewiring probability $p > 0$, and the
transition temperature $T_c$ increases as $p$ is raised.
A similar mean-field-type  phase transition was found in small-world 
networks generated from 2D and 3D regular lattices \cite{he02a,ha03},
as well as for the $XY$ model in networks generated from one-dimensional 
chains \cite{ki01}.
In recent years, the Ising model has been thoroughly studied in
complex networks, such as the so-called scale-free networks, where
several unusual features were observed \cite{le02,do02b,ig02,he04}.

Here we study the antiferromagnetic (AFM) Ising model in small-world 
networks generated by rewiring a 2D square lattice. 
One expects that the AFM ordering present in the regular lattice at low
temperature will be lost when random connections are introduced,
for an increasing number of bonds will be frustrated as $p$ rises. 
In particular, this model includes the two basic ingredients
necessary to have a spin glass (SG), namely, disorder and frustration. 
The former appears due to the random long-range connections introduced 
in the rewiring process, and the latter because half of these rewired links
connect sites located in the same sublattice of the starting regular
lattice.  

In some spin-glass models, such as the Sherrington-Kirkpatrick model, 
all spins are mutually connected \cite{fi91,my93}.
An intermediate step between these globally connected networks and
finite-dimensional models consists in studying spin glasses on random
graphs with finite (low) connectivity \cite{ka87,de01,bo03,ki05}.
A further step between random graphs with finite mean connectivity
and regular lattices is provided by small-world networks, where one
can modify the degree of disorder by changing the rewiring 
probability $p$.
Then, for the AFM Ising model on small-world networks, we expect 
to find features close to those of short-range spin-glass systems. 
In this line, a spin-glass phase has been recently found and characterized 
for the AFM Ising model in scale-free networks \cite{ba06}.
In this paper, we employ Monte Carlo (MC) simulations to study the 
paramagnetic to spin-glass phase transition occurring in small-world
networks. Apart from temperature and system size, another variable is
the rewiring probability, which controls the degree of disorder,
and allows us to interpolate from a paramagnetic-AFM transition
at $p=0$ to a paramagnetic-SG transition in a random graph at $p=1$. 
 
The paper is organized as follows.
In Sec.~II we describe the networks and the computational method
employed here. 
In Sec.\,III we give results for the heat capacity, energy, and 
spin correlation, as derived from MC simulations. 
In Sec.\,IV we present and discuss the overlap parameter, transition
temperature, and absence of long-range ordering.
The paper closes with the conclusions in Sec.\,V.

\section{Model and method}

We consider the Hamiltonian:
\begin{equation}
H = \sum_{i < j} J_{ij} S_i S_j   \, ,
\end{equation}
where $S_i = \pm 1$ ($i = 1, ..., N$), and the coupling matrix
$J_{ij}$ is given by
\begin{equation} 
J_{ij}  \equiv \left\{
     \begin{array}{ll}
         J (> 0) & \mbox{if $i$ and $j$ are connected,} \\
         0   & \mbox{otherwise.}
     \end{array}
\right.
\label{Jij}
\end{equation}    
This means that each edge in the network is an AFM interaction between 
spins on the two linked nodes.
Note that, contrary to the usually studied models for spin glasses,
in this model all couplings are antiferromagnetic.
This model with AFM couplings can be mapped onto one in which all
unrewired bonds are ferromagnetic (FM), and the rewired links are 50\%
FM and 50\% AFM.  In the limit $p \to 0$, this mapping is the well-known
correspondence between AFM and FM Ising models on bipartite lattices
\cite{la99}.
In the limit $p \to 1$, our AFM Ising model is equivalent to a spin-glass
model on a random graph of mean connectivity $\langle k \rangle = 4$, 
with 50\% AFM and 50\% FM bonds. 

Small-world networks have been built up according to the model
of Watts and Strogatz \cite{wa98,wa99}, i.e.,
we considered in turn each of the bonds in the starting 2D lattice and 
replaced it with a given probability $p$ by a new connection.
In this rewiring process, one end of the selected bond is changed
to a new node chosen at random in the whole network.
We impose the conditions: (i) no two nodes can have more than
one bond connecting them, and (ii) no node can be connected by a link
to itself.  With this procedure we obtained networks where 
more than 99.9\% of the sites were connected in a single component.
Moreover, this rewiring method keeps constant the total number of links 
in the rewired networks, and the average connectivity  $\langle k \rangle$
coincides with $z = 4$. This allows us to study the effect of disorder
upon the physical properties of the model, without changing the mean
connectivity.

We note that other ways of generating small-world networks from regular
lattices have been proposed \cite{ne99,ne00b}. In particular, instead of 
rewiring each bond with probability $p$, one can add shortcuts between
pairs of sites taken at random, without removing bonds
from the regular lattice. This procedure turns out to be more convenient
for analytical calculations, but does not keep constant the mean
connectivity $\langle k \rangle$, which in this case increases with $p$.
Spin glasses on such small-world networks, generated from a one-dimensional
ring, have been studied earlier by replica symmetry breaking \cite{we05} and
transfer matrix analysis \cite{ni04}.

From the 2D square lattice, we generated small-world networks of different
sizes.  The largest networks employed here included 200 $\times$ 200 nodes. 
Periodic boundary conditions were assumed. 
For a given network, we carried out Monte Carlo simulations at several
temperatures, sampling the spin configuration space by the Metropolis update 
algorithm \cite{bi97}, and using a simulated annealing procedure. 
Several variables characterizing the considered
model have been calculated and averaged for different values of $p$, $T$,
and system size $N$.
In general, we have considered 300 networks for each rewiring probability
$p$, but we used 1000 networks to determine accurately the transition 
temperature from paramagnetic to SG phase.
In the following, we will use the notation $\langle ... \rangle$ to
indicate a thermal average for a network, and $[ ... ]$ for an average
over networks with a given degree of disorder $p$.

\section{Thermodynamic observables}

The heat capacity per site, $c_v$, was obtained from the energy fluctuations
$\Delta E$ at a given temperature, by using the expression
\begin{equation} 
c_v = \frac {[ (\Delta E)^2 ]} {N T^2}  \,  ,
\end{equation} 
where $(\Delta E)^2 = \langle E^2 \rangle - \langle E \rangle^2$.
We have checked that the results coincide within numerical noise with
those derived by calculating $c_v$ as $[d \langle E \rangle / d T] / N$.
Note that we take the Boltzmann constant $k_B = 1$.

\begin{figure}
\vspace{-2.0cm}
\includegraphics[width= 9cm]{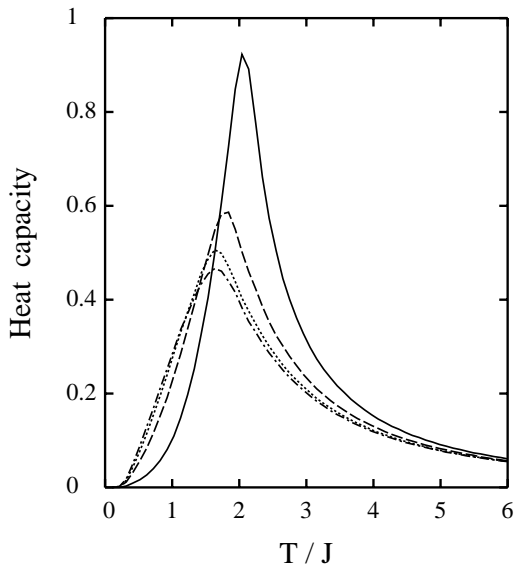}
\vspace{-2.5cm}
 \caption{
Heat capacity per site $c_v$ vs temperature for small-world
networks generated from a 2D lattice of size $80 \times 80$.
The plotted curves correspond to different values of the rewiring
probability $p$. From top to bottom: $p$ = 0.1, 0.3, 0.5, and 1.
} \label{fig1} \end{figure}

The temperature dependence of $c_v$ is displayed in Fig.~\ref{fig1} for
saveral values of the rewiring probability $p$ and for networks built 
up from a $80 \times 80$ 2D lattice. For increasing $p$, one observes 
two main features: the maximum of $c_v$ shifts to lower $T$ and the
peak broadens appreciably. This broadening agrees with the behavior
expected for systems with increasing disorder, similarly to that found
for the FM Ising model in these networks \cite{he02a}. However, in the
AFM model, the shift of the peak to lower temperature suggests a phase 
transition with a temperature $T_c$ that decreases as $p$ is raised, 
contrary to the FM case, where an increase in $T_c$ with $p$ was 
observed. This difference between both Ising models on small-world
networks occurs in addition to the nature of the transition itself,
which in the FM case is a paramagnetic-ferromagnetic transition vs 
a paramagnetic-SG transition in the AFM model (see below).
A decrease in $T_c$ for the AFM Ising model in this kind of networks was
also suggested in Ref. \onlinecite{ca04} from the behavior of the
heat capacity for several values of $p$.

The increase in disorder as $p$ is raised is
accompanied by an increase in frustration of the links 
at low temperatures. This can be quantified by the low-temperature
energy of the system, which will rise as the rewiring probability
is raised. 
To obtain insight into this energy change for $p$ near zero
(low disorder), let us remember that
the square lattice is bipartite, in the sense that one can
define two alternating sublattices, say A and B, so that
neighbors of each node in sublattice A belong to sublattice B,
and vice versa. In the rewiring process, one introduces links
between nodes in the same sublattice, and the resulting networks
are no longer bipartite.
However, for small rewiring probability $p$, we can still speak about
two sublattices, with some ``wrong'' connections.
Since each link is rewired with probability $p$, each connection in the
starting regular lattice will be transformed into a wrong connection
(of types A--A or B--B) with probability $p/2$.
The remaining links are of A--B type, and
the number of wrong connections is on average $z N p / 4$.
Then, for small $p$, the lowest energy can be approximated by 
$E_{\text m} = - z N J (1 - p)/2$, under the assumption that the
AFM long-range ordering of the square lattice is still preserved.
For z = 4, we have an energy per node: $e_{\text m} = - 2 (1-p) J$.
Note that for finite $p$ the low-temperature long-range ordering in fact 
decays due to the appearance of domains driven by the rewired connections
(see below), but the AFM ordering is a good reference to obtain insight
into the energy change as a function of rewiring probability $p$.

\begin{figure}
\vspace{-2.0cm}
\includegraphics[width= 9cm]{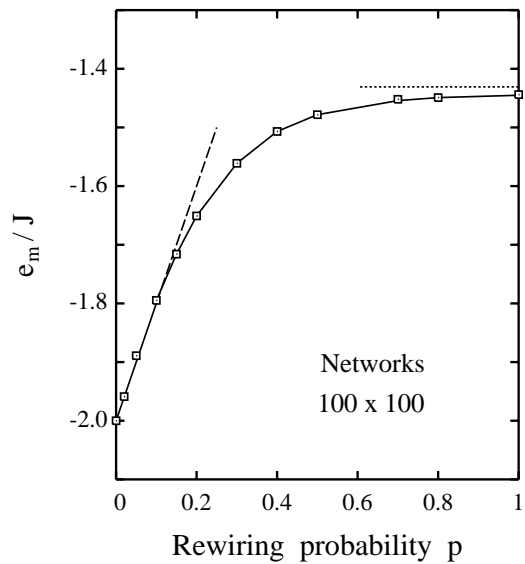}
\vspace{-2.5cm}
\caption{
Minimum energy per site obtained in our simulations for the AFM Ising
model on small-world networks with $N = 10^4$ nodes.
The dashed line corresponds to $e_{\text m} = (-2 + 2p) J$, as
explained in the text.
The dotted line indicates the ground-state energy obtained in
Ref.~\onlinecite{bo03} for a spin glass on random networks with
$\langle k \rangle = 4$.
The solid line is a guide to the eye.
} \label{fig2} \end{figure}

We now turn to the results for the minimum energy reached in 
our simulations for different $p$ values, which are shown in
Fig.~\ref{fig2}. 
For rising $p$, $E_{\text m}$ increases from the
value corresponding to AFM ordering in the regular lattice, 
$e_{\text m} = - z J / 2$. 
The dashed line in Fig.~\ref{fig2} displays the behavior expected
for small $p$, in the case of a strict AFM ordering on the underlaying
lattice. This estimation is close to 
the minimum energy obtained in our simulations for $p<0.15$.
For larger $p$ values, it departs appreciably from the results of the
simulations, and $e_{\text m}$ lies below the dashed line. 
In the limit $p=1$ we find a value $e_{\text m} = -1.444(2) J$.
In this limit, our small-world networks are very close to random networks
with a Poisson distribution of connectivities, but are not identical to
the latter because of the restriction that no nodes have zero links,
imposed in the rewiring process \cite{ba00,he02b}. 
For a $\pm J$ Ising spin glass on random graphs with a Poisson distribution 
of connectivities and $\langle k \rangle = 4$, 
Boettcher \cite{bo03} found a ground-state
energy $e_{\text m} = -1.431(1) J$ by using extremal optimization.
This value is plotted in Fig.~\ref{fig2} as a dotted line close to $p=1$, 
and is
near the minimum energy we found for the small-world networks in this limit.
For a more direct comparison with random networks, we have carried out
some simulations for small-world networks with $p=1$, where we allowed 
the presence 
of isolated sites (with connectivity $k=0$). For the AFM Ising model in 
these networks, we found a minimum energy $e_{\text m} = -1.438(2) J$, 
between those of our standard networks (with minimum connectivity $k=1$) and
random networks in Ref.~\onlinecite{bo03}. Note that our error bar in
$e_{\text m}$ corresponds to a standard deviation in the distribution 
of minimum energy obtained for different networks. 
We emphasize that the energy $e_{\text m}$ found here for each value of
$p$ is an upper limit for the lowest energy of the system. 

\begin{figure}
\vspace{-2.0cm}
\includegraphics[width= 9cm]{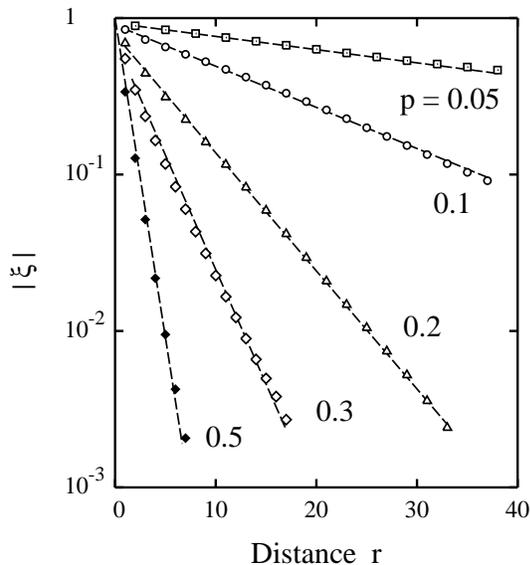}
\vspace{-2.5cm}
\caption{
Absolute value of the spin correlation function vs distance in
small-world networks with $N = 10^4$ nodes, at temperature $T = 1.5 J$.
The dimensionless distance between nodes, $r = d/d_0$, is measured
on the starting regular lattice. Data are shown for several values
of $p$, as indicated by the labels.
} \label{fig3} \end{figure}

Even though the random connections present in small-world networks
introduce disorder in the starting regular lattice, these networks still
keep memory of the original bipartite lattice, but the actual meaning of the 
partition in two sublattices is gradually reduced as $p$ rises.
This can be measured by the number of wrong links, which amounts to 
a fraction $p/2$ of the total number of links, as indicated above.
In the limit $p=1$, half of the links connect sites in the same original
sublattice, and the memory of the partition in sublattices has completely 
disappeared.
One can visualize the loss of AFM ordering on the 2D lattice, by plotting
the spin correlation vs distance for several values of $p$.
We define $\xi$ as
\begin{equation}
 \xi(r) =  [ \langle S_i S_j \rangle_r ]   \; , 
\label{xi}
\end{equation}
where the subscript $r$ indicates that the average is taken for the 
ensemble of pairs of sites at distance $r$.
Note that $r = d/d_0$ refers here to the dimensionless distance between 
sites in the starting regular lattice, not to the actual topological distance 
or minimum number of links between nodes in the rewired networks
($d_0$ is the distance between nearest neighbors).
The correlation $\xi(r)$ is shown in Fig.~\ref{fig3} for several values
of the rewiring probability $p$, at temperature $T = 1.5 J$.
This temperature is below the critical temperature $T_c$ of the
paramagnetic-SG transition for all values of $p$ (see below).
As expected, $\xi(r)$ decreases faster for larger $p$, and
vanishes at $p=1$ for any distance $r \ge 1$. In general, after a short
transient for small $r$, we find an exponential decrease of the spin
correlation with the distance.
This indicates that, in spite of the disorder present in the networks
for $p>0$, there remains some degree of short-range AFM ordering on the 
starting regular lattice, which is totally lost in the limit $p \to 1$.

\section{Spin-glass behavior}

\subsection{Overlap parameter}

As is usual in the study of spin glasses, we now consider two copies of
the same network, with a given realization of the disorder, and study
the evolution of both spin systems with different initial values
of the spins and different random numbers for generating the spin 
flips \cite{pa83,ka96}.
It is particularly relevant the overlap $q$ between the two copies,
defined as
\begin{equation}
q =  \frac{1}{N}\sum_{i} S_i^{(1)} S_i^{(2)},
\label{qq}
\end{equation}
where the superscripts (1) and (2) denote the copies. Obviously, $q$
is defined in the interval $[-1,1]$.

We have calculated the overlap parameter $q$ for small-world networks with
various rewiring probabilities $p$, and obtained its probability
distribution
$P(q)$ from MC simulations.
This distribution is shown in Fig.~\ref{fig4} for $p$ = 0.1 and 0.5 at
several temperatures.
At high temperature, $P(q)$ shows a single peak centered at
$q=0$, characteristic of a paramagnetic state. The width of this
peak is a typical finite-size effect, which should collapse to
a Dirac $\delta$ function at $q=0$ in the thermodynamic limit
$N \to \infty$.
When the temperature is lowered, the distribution $P(q)$ broadens,
as a consequence of the appearance of an increasing number of edges
displaying frustration.
At still lower temperatures, two peaks develop in $P(q)$, symmetric
respect to $p=0$, and characteristic of a spin-glass phase
\cite{ka96,ka02,ba06}.

\begin{figure}
%\vspace*{-1.5cm}
\vspace{-2.0cm}
\includegraphics[width= 9cm]{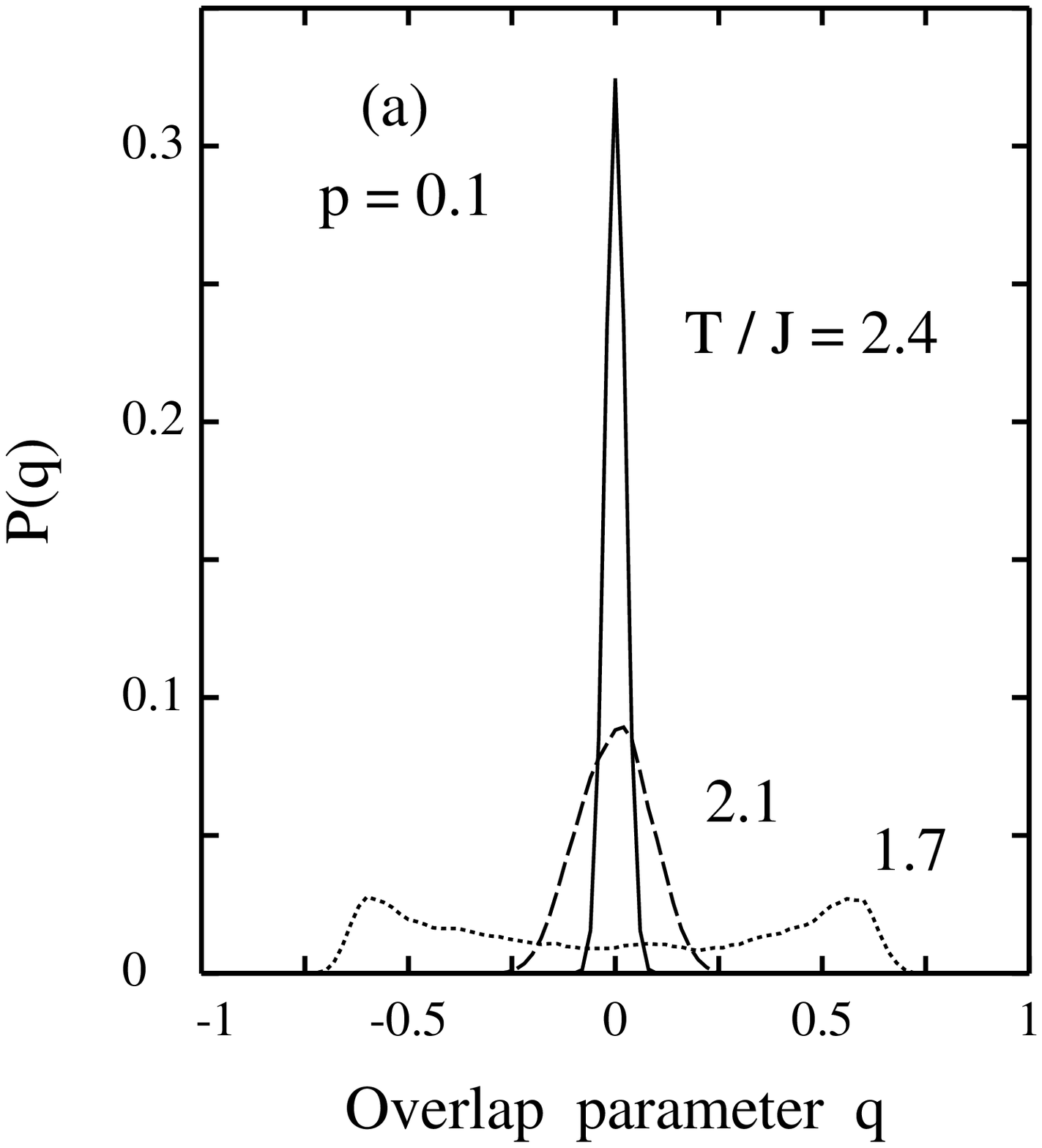}
%\vspace{-2.5cm}
\end{figure}
                                                                                
\begin{figure}
\vspace{-4.0cm}
%\vspace{-2.0cm}
\includegraphics[width= 9cm]{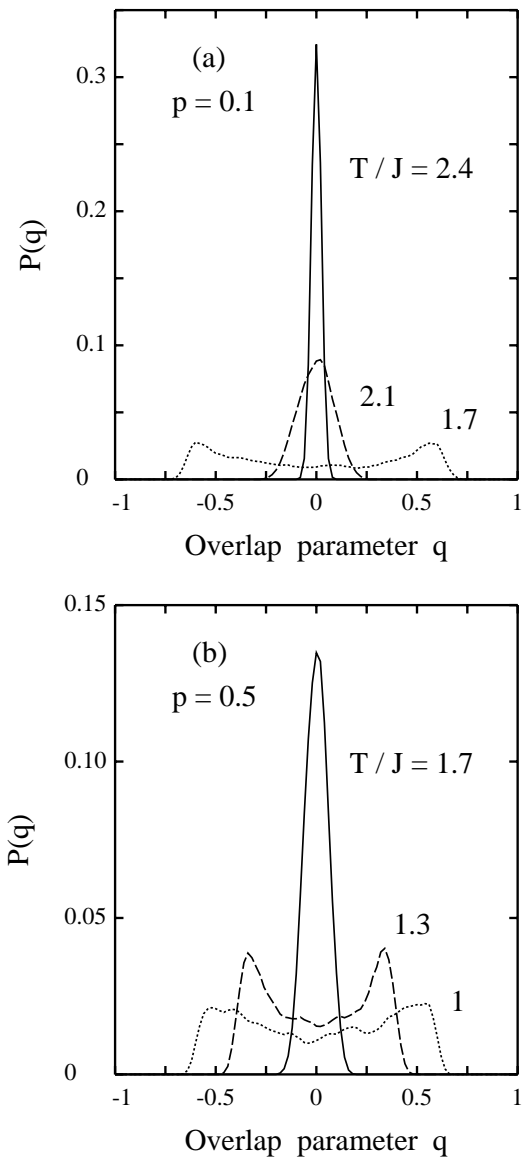}
\vspace{-2.5cm}
\caption{
Distribution of the overlap parameter $q$ for two rewiring probabilities
and various temperatures, as derived from our simulations for networks
with $N = 10^4$ nodes.
(a) $p = 0.1$ at $T/J$ = 2.4, 2.1, and 1.7;
(b) $p = 0.5$ at $T/J$ = 1.7, 1.3, and 1.
} \label{fig4} \end{figure}

\begin{figure}
\vspace{-2.0cm}
\includegraphics[width= 9cm]{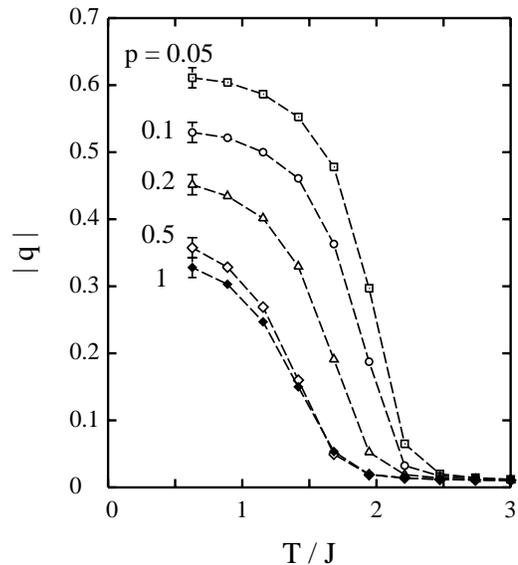}
\vspace{-2.5cm}
\caption{
Average of the absolute value of the overlap parameter
$[ \langle q \rangle ]$ for various rewiring probabilities $p$ and
several temperatures.
Symbols are data points derived from MC simulations on networks of size
$N = 10^4$.
} \label{fig5} \end{figure}

Information on the ``freezing'' of the spins as temperature is
lowered can be obtained from the evolution of the average value of 
$|q|$, for a given degree of disorder $p$.
This average value is shown in Fig.~\ref{fig5} as a function
of temperature for several rewiring probabilities $p$.
It is close to zero at the high-temperature
paramagnetic phase, and increases as temperature is reduced,
indicating a break of ergodicity associated to the spin-glass
phase \cite{ba06,ka96}. 
For $p=0$, $|q|$ converges to unity at low temperatures, reflecting 
the AFM ordering present in the regular lattice. For increasing
$p$, we find a decrease in the low temperature $|q|$ values, 
due to an increasing degree of frustration.

\subsection{Transition temperature}

\begin{figure}
\vspace{-1.5cm}
\includegraphics[width= 9cm]{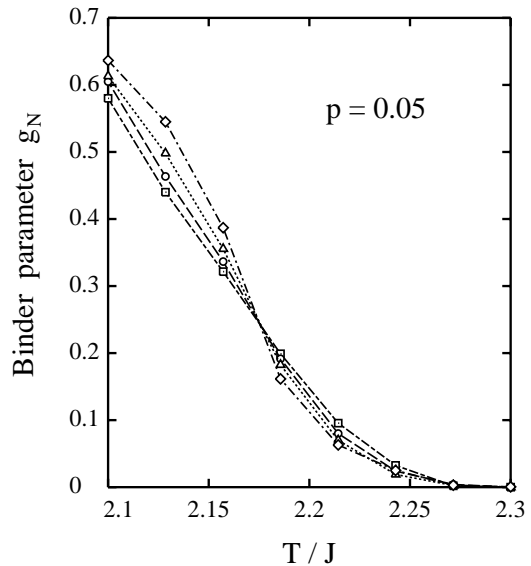}
\vspace{-2.5cm}
\caption{
Fourth-order Binder's cumulant $g_N$ as a function of temperature
for small-world
networks generated from 2D square lattices with rewiring probability
$p = 0.05$. Symbols represent different system sizes $N = L^2$: squares,
$L = 60$; circles, $L = 80$; triangles, $L = 100$; diamonds, $L = 150$.
Error bars are on the order of the symbol size.
} \label{fig6} \end{figure}

The overlap parameter $q$ can be used to obtain accurate values of 
the paramagnetic-SG transition temperature, by using the fourth-order 
Binder cumulant \cite{bi97,ka96} 
\begin{equation} \label{Binder}
g_N(T) = \frac{1}{2}\left(3 - \frac{ \left[\langle q^4\rangle\right]_N }
   {\left[ \langle q^2 \rangle \right]^{2}_N} \right) \, .
\end{equation}
This parameter can change in the interval [0,1]. 
One has $g_N =0$ for a Gaussian
distribution $P(q)$ (high temperatures), and $g_N = 1$ when 
$|q| = 1$ (in the particular case of a single ground state).
In general $g_N$ rises for decreasing temperature, and $T_c$
can be obtained from the crossing point for different network sizes $N$. 
As an example, we present in Fig.~\ref{fig6} $g_N(T)$ as a function of
temperature for several system sizes and a rewiring probability 
$p$ = 0.05. From the crossing point we find $T_c/J = 2.175 \pm 0.005$
for this value of $p$.

\begin{figure}
\vspace{-2.0cm}
\includegraphics[width= 9cm]{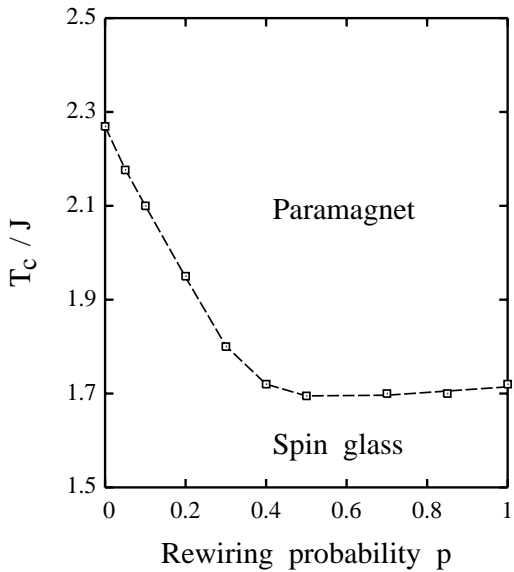}
\vspace{-2.5cm}
\caption{
Transition temperature $T_c$ as a function of the rewiring probability
$p$ for small-world networks. Error bars are on the order of the symbol
size.
The dashed line is a guide to the eye.
} \label{fig7} \end{figure}

By using this procedure, we have calculated the transition temperature
$T_c$ for several values of $p$, and the results so obtained are shown 
in Fig.~\ref{fig7}.
For small $p$, $T_c$ decreases linearly from the transition temperature
corresponding to the AFM model on the 2D square lattice, and for $p > 0.4$
it saturates to a value of about $1.7 J$.   
Close to $p$ = 0, we find a change in $T_c$ induced by the 
long-range  links: $T_c = T_c^0 - a p J$, where $T_c^0$ is the
paramagnetic-AFM transition temperature in the square lattice and 
$a \approx 2$. 

It is interesting the approximately linear decrease in $T_c$ for increasing
$p$ up to $p \approx 0.3$. This decrease could be expected from the
larger number of frustrated links appearing as $p$ is raised. For increasing
frustration, the paramagnetic phase is favored, and the spin glass appears
at lower temperature ($T_c$ is reduced). This change of $T_c$ as a function
of rewiring probability could be also expected from the behavior of the heat
capacity shown in Fig.~\ref{fig1}. For $p > 0.4$, small-world networks
behave in this respect similarly to Poissonian random networks, in the sense 
that the transition temperature is roughly independent of $p$, and is close
to that found for the strong-disorder limit $p = 1$.

\subsection{Absence of long-range AFM ordering}

Even though all data indicate that the AFM Ising model in small-world 
networks yields a spin-glass phase at low temperature, one can
ask if such disordered phase appears for any finite value of the
rewiring probability. One could argue that some residual long-range
AFM ordering could be present for finite but small $p$ values.
From our considerations in the preceding
sections, one could think that, for small $p$, the low-temperature
phase still keeps the long-range ordering characteristic of the
2D regular lattice, with some defects caused by the long-range
connections. 
To analyze this question, we consider a staggered magnetization defined 
for the square lattice as usual:
\begin{equation}
  M_s = M_A - M_B  \, ,
\end{equation}
with
\begin{equation}
  M_A = \sum_{i \in A} S_i  \, ,
\end{equation}
and similarly for $M_B$. Our question then refers to the possibility
of a finite value for $M_s$ for small-world networks with $p>0$.

\begin{figure}
\vspace{-2.0cm}
\includegraphics[width= 9cm]{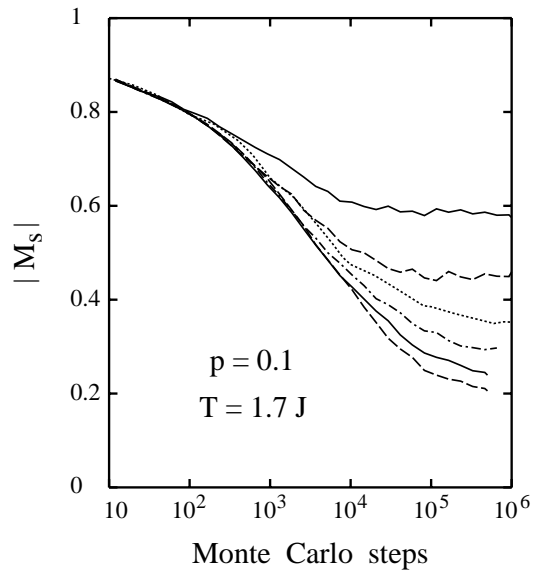}
\vspace{-2.5cm}
\caption{
Relaxation of the staggered magnetization on the underlying regular lattice
for small-world networks with $p = 0.1$, from simulations starting from an
ordered AFM configuration. The plot shows the evolution of
$|M_s|$ at $T = 1.7 J$ and different system sizes
$L \times L$. From top to bottom: $L$ = 40, 60, 80, 100, 150, and 200.
} \label{fig8} \end{figure}

To check this point, we have carried out simulations starting from an 
AFM ordered configuration and followed the evolution of $M_s$.
We prefer this procedure to directly calculating the low-temperature 
staggered magnetization from simulated annealing, since in this case a
long-range ordering can be difficult to find for large nerworks, due to
the appearance of different spin domains.
Thus, we analyzed the decay of $M_s$ at temperatures lower than the
transition temperature $T_c$ for different system sizes. 
In particular, in Fig. 8 we show the relaxation of $|M_s|$ on networks
with $p = 0.1$ at a temperature $T = 1.7 J$, well below the transition 
temperature for this rewiring probability ($T_c/J = 2.10 \pm 0.01$).
For each system size, $|M_s|$ decreases from unity to reach a plateau
at a finite value, which is clearly a finite-size effect, as seen in the
figure.  As the system size increases, such a plateau appears after longer
simulation times, and the corresponding value of $|M_s|$ decreases.
These results are consistent with the decay of the spin correlation $\xi(r)$
at temperatures below $T_c$, as presented in Fig.~\ref{fig3}  for several values of
$p$.

For $p < 0.1$, a relaxation of $M_s$ is expected to appear for larger
system sizes and longer simulation runs. Everything indicates that
at low $T$ the long-range ordering disappears in the thermodynamic limit
$N \to \infty$ for any $p > 0$.
This is in line with earlier results for the FM Ising model on this kind
of networks, in the sense that the paramagnetic-FM transition occurring in
those systems changes from an Ising-type transition at $p = 0$ to a 
mean-field-type one (typical of random networks) for any finite value of 
the rewiring probability $p > 0$ \cite{ba00}.
The observation of this mean-field character for the paramagnetic-FM
transition requires system sizes that increase as the rewiring probability
is lowered \cite{he02a}, similarly to the decay of the staggered 
magnetization in the AFM case shown here.

\section{Conclusions}

The combination of disorder and frustration in the AFM Ising model on
small-world networks gives rise to a spin-glass phase at low
temperatures.
The transition temperature from a high-temperature paramagnet 
to a low-temperature spin-glass phase goes down for increasing
disorder, and saturates to a value $T_c \approx 1.7 J$ for $p > 0.4$.

The overlap parameter provides us with clear evidence of the frustration
associated to the spin-glass phase at low temperatures. The degree of 
frustration increases as the disorder (or rewiring probability) rises.
For small rewiring probability $p$, the energy of the ground state increases 
linearly with $p$ up to $p \approx 0.15$, and for larger $p$, it converges
to $e_m = - 1.44 J$.
In the limit $p \to 1$ one recovers the behavior of a $\pm J$ Ising spin
glass in random networks.

An interesting feature of the physics here is that a small fraction
of random connections is able to break the long-range AFM ordering present
in the 2D square lattice at low temperature.

\begin{acknowledgments}
The author benefited from useful discussions with M. A. Ramos.
This work was supported by Ministerio de Educaci\'on y
Ciencia (Spain) under Contract No. FIS2006-12117-C04-03.  \\
\end{acknowledgments}

\end{document}